\documentclass[prl,twocolumn,superscriptaddress]{revtex4}
\usepackage{amsmath, amssymb,graphicx,color,bm,epstopdf}
\usepackage[dvipsnames]{xcolor}
\usepackage{bbold}
\usepackage{braket}
\usepackage{comment}

\begin{document}

\title{Quantum Time Crystals from Hamiltonians with Long-Range Interactions}

\author{Valerii K. Kozin}
\affiliation{Science Institute, University of Iceland, Dunhagi 3, IS-107, Reykjavik, Iceland}
\affiliation{ITMO University, Kronverkskiy prospekt 49, Saint Petersburg 197101, Russia}

\author{Oleksandr Kyriienko}
\affiliation{Department of Physics and Astronomy, University of Exeter, Stocker Road, Exeter EX4 4QL, UK}
\affiliation{Department of Nanophotonics and Metamaterials, ITMO University, St. Petersburg, 197101, Russia}
\affiliation{NORDITA, KTH Royal Institute of Technology and Stockholm University, Roslagstullsbacken 23, SE-106 91 Stockholm, Sweden}

\begin{abstract}
Time crystals correspond to a phase of matter where time-translational symmetry (TTS) is broken. Up to date, they are well studied in open quantum systems, where external drive allows to break discrete TTS, ultimately leading to Floquet time crystals. At the same time, genuine time crystals for closed quantum systems are believed to be impossible. In this study we propose a form of a Hamiltonian for which the unitary dynamics exhibits the time crystalline behavior and breaks continuous TTS. This is based on spin-1/2 many-body Hamiltonian which has long-range multispin interactions in the form of spin strings, thus bypassing previously known no-go theorems. We show that quantum time crystals are stable to local perturbations at zero temperature. Finally, we reveal the intrinsic connection between continuous and discrete TTS, thus linking the two realms.
\end{abstract}

\maketitle

\textit{Introduction.---}Crystalline structures are ubiquitous in nature, and surround us in everyday life. Formally, a crystal is defined as a structure of atoms which is periodically ordered in space, and is stable to external perturbations \cite{AshcroftMermin}. Generalized to various other constituents, the concept was found applicable in many areas of science, including photonic crystals in optics \cite{Lodahl2015,Istrate2006}, synthetic cold atom lattices \cite{Bloch2008}, granular materials \cite{Beloborodov2007} etc. 
While thought to be inherent to the spacial degrees of freedom, the contemporary understanding of physics in \emph{space-time} posed the question of possibility to have a crystalline structure in the time dimension. This would correspond to a time-periodic behavior, which persists perpetually. For the classical case, the time crystal (TC) was described by Shapere and Wilczek \cite{Shapere2012}. Finally, the quantum case was postulated by Frank Wilczek \cite{Wilczek2012}, theorizing the breaking of time-translational symmetry. This was suggested to be observed in superconducting rings \cite{Wilczek2013} or trapped ion \cite{Li2012} systems. However, despite the basic time crystal formulation remains an intriguing concept, the subsequent studies have shown that ground state TC cannot be realized in the simple setting \cite{Bruno2013C,Bruno2013}, and are ruled out by the no-go theorem.

The formalized definition of a quantum time crystal was set by Watanabe and Oshikawa (W-O) \cite{Oshikawa}, who proposed to closely follow the definition for spatial crystals. It should be noted, however, that there is a difference between breaking the space and time-translational symmetries. In the case of spontaneous breaking of the space translational symmetry the ground state is highly degenerate in the thermodynamic limit. This results in the fact that the ground state of a crystal does not have a definite momentum even though the Hamiltonian commutes with the momentum operator. The continuous time translational symmetry is present in isolated systems where the Hamiltonian does not depend on time (the Hamiltonian commutes with the evolution operator) but the ground state must have a certain energy by definition. Instead, W-O defined the condition for the correlation function to be a periodic function of time $t$ as
\begin{align}
\label{eq:WO_definition}
    \lim\limits_{N\rightarrow \infty} \langle \hat{\Phi}(t) \hat{\Phi}(0) \rangle = f(t),
\end{align}
where the operator $\hat{\Phi}(t)$ corresponds to an integrated order parameter (e.g. average density or magnetization) at time $t$, and the averaging over the ground state of the system is considered in the thermodynamic limit of infinite number of lattice sites $N$. At the same time, W-O have proven that non-vanishing periodic oscillations are impossible unless two conditions are violated: 1) the system is out-of-equilibrium; or 2) the interaction is of long-range type, decaying slower than power law $\sim r^{-\alpha}$ for $\alpha>0$. This left the question of possibility for quantum time crystals to remain in the air. 

Analyzing the first constraint, the consideration of the driven quantum system previously has allowed breaking TTS in \emph{discrete} time \cite{Sacha2015}. The idea relies on the successive application of driver and interaction unitaries, and ultimately led to the appearance of Floquet or discrete time crystals (DTC) \cite{Else2016,Khemani2016,Sacha2015}. For the chain of spin-1/2 particles, the protocol reads as simultaneous spin flip application to the entire chain, which leads to the oscillations of magnetization, followed by Ising type interaction unitary. The major feature of the described system is a subharmonic response which is stable to drive perturbations \cite{Else2016,Yao2017}.  As the driven system is prone to heating in high frequency limit, it also involves many-body localized unitary evolution, although later this condition was relaxed if the finite time scale is considered \cite{Else2017,Huang2018,YuAngelakis2019}. The successful theoretical prediction of time-periodic structures in driven systems was followed by its experimental observation in trapped ions \cite{ZhangMonroe2017}, silicon vacancy-based quantum simulators \cite{ChoiLukin2017}, superfluid helium-3 \cite{AuttiVolovik2018}, NMR systems \cite{Pune2018,Rovny2018}. Finally, many works targeting other platforms and generalizations were presented, including TC in fermionic lattices \cite{Huang2018} and Majorana fermions \cite{NTU2019}, dissipative DTCs \cite{OSullivan2018}, topological DTCs \cite{Giergiel2019}, discrete time quasicrystals \cite{Giergiel2019b}, cosmological space-time crystals \cite{DasGhosh2018,DasGhosh2019}, and described the dynamical phase transition for DTCs \cite{Kosior2018}. The emergent field up to date was reviewed in several in-depth studies \cite{Sacha2018,Else2019}. Further directions represent exploration of quantum scar physics in nonequilibrium setting \cite{Turner2018,WWHo2019,Choi2019,Iadecola2019} or modification of TC definition to nonlocal order parameters \cite{Medenjak2019}. Finally, to study the breaking of continuous TTS, several open system realizations were considered up to date, including fermionic cold atom lattices~\cite{Buca2019} and Rydberg atoms~\cite{Gambetta2019}.

Although the number of systems demonstrating discrete time crystalline behavior grows rapidly for driven-dissipative and Floquet systems, the original issue of possibility to break \emph{continuous} time-translational symmetry and existence of genuine time crystals for closed quantum system remains an open challenge. In this Letter we use the original time crystal definition and exploit the second loophole in W-O no-go theorem to show that there is a class of static Hamiltonians which host TC phase for the unitary dynamics. In particular, using W-O definition for spin-1/2 many-body system, we find Hamiltonian with spin string-type interactions which corresponds to a system with a ground state being a nondegenerate maximally entangled state. The minimal size of the spin-string to support TC phase is $N/2$ for a system of $N$ particles, thus formally bypassing the applicability range of W-O theorem \cite{Oshikawa}. The stability of the system is studied, showing resilience to local perturbations. 

\textit{Continuous time crystal Hamiltonian.---}We start by postulating that certain many-body Hamiltonians which contain terms involving all particles can support time crystalline phase and exhibits spontaneously broken continuous TTS. Here, we present a method of constructing a Hamiltonian satisfying the W-O definition of a quantum time crystal. To be concrete, we focus on an $N$-qubit system, and as an operator $\hat{\Phi}$ in the definition \eqref{eq:WO_definition} we consider the total magnetization $\hat{M}_{z}= \sum_{i=1}^{N} \sigma^{(i)}_z/N$ along the z-axis, where $\sigma^{(i)}_{x,y,z}$ denote Pauli matrices acting on qubits at site $i$. In this case, the Watanabe-Oshikawa TC definition in Eq.~\eqref{eq:WO_definition} recasts as $\lim\limits_{N\rightarrow\infty}\langle \hat{M}_{z}(t)\hat{M}_{z}(0)\rangle=f(t)$ where $f(t)$ must be periodic. To construct a Hamiltonian satisfying this definition we start with a state $\lvert E_0\rangle$ from the Hilbert space that we chose as a ground state at energy $\epsilon_0$. Acting on $\lvert E_0\rangle$ by a non-unitary operator $\hat{M}_{z}$ and renormalizing the vector gives the partner state $\lvert E_1\rangle=\hat{M}_{z}\lvert E_0\rangle/ \sqrt{\langle E_0\rvert \hat{M}_{z}^2\lvert E_0\rangle}$. The crucial point here is to demand that $\lvert E_0\rangle$ is chosen such that $\lvert E_1\rangle$ is orthogonal to $\lvert E_0\rangle$, i.e. $\langle E_0\lvert \hat{M}_{z}\rvert E_0\rangle=0$. Having done that, we assign an energy $\epsilon_1 > \epsilon_0$ to the state $\lvert E_1\rangle$, and construct the Hamiltonian using projectors for the many-body states as
\begin{equation}\label{eq:ham_gen}
    \hat{H} = \epsilon_0 \lvert E_0\rangle\langle E_0\rvert + \epsilon_1\lvert E_1\rangle\langle E_1\rvert + \sum_{j>1} \epsilon_j\lvert E_j\rangle\langle E_j\rvert,
\end{equation}
where arbitrary real numbers $\epsilon_1, \epsilon_2, ... ,\epsilon_{2^N-1}$ must all be greater than $\epsilon_0$, with $\epsilon_0$ being a ground state energy, and arbitrary state vectors are chosen such that $\langle E_{0,1}\rvert E_j\rangle=0$ for $j>1$. We claim that Eq.~\eqref{eq:ham_gen} describes the quantum time crystal Hamiltonian if certain restrictions on $\lvert E_0\rangle$ are respected. To find these restrictions we calculate the correlation function at finite size $N$ of the system given in Eq.~\eqref{eq:WO_definition} for the Hamiltonian~\eqref{eq:ham_gen}, which yields
\begin{align}
    &f_N(t)=\langle E_0\rvert \hat{M}_{z}(t)\hat{M}_{z}(0)\rvert E_0 \rangle=\langle E_0\rvert e^{i\hat{H}t}\hat{M}_{z} e^{-i\hat{H}t}\hat{M}_{z}\rvert E_0 \rangle \nonumber\\
    &=e^{i \epsilon_0 t} \langle E_0\rvert \hat{M}_{z}^2\lvert E_0\rangle \langle E_1\rvert e^{-i\hat{H}t}\rvert E_1 \rangle= \mathcal{O} e^{-i (\epsilon_1-\epsilon_0) t},
\end{align}
where $\mathcal{O} = \langle E_0\rvert \hat{M}_{z}^2\lvert E_0\rangle$ denotes the order parameter. If it remains non-zero $\mathcal{O} \neq 0$ at $N\rightarrow\infty$ (as for the case of long-range order in crystals \cite{Oshikawa}) and the energy difference $\omega=\epsilon_1-\epsilon_0 \neq 0$, then $\lim_{N\rightarrow\infty}f_N(t)=f(t)$ is a periodic function of time in the thermodynamic limit, oscillating at frequency $\omega$. The latter condition on $\omega$ can be fulfilled easily by construction, whereas the former one needs an additional consideration. 

We remind that our goal is to find such a state $\lvert E_0\rangle$ that: (a) $\langle E_0\rvert \hat{M}_{z}\lvert E_0\rangle=0$ for any $N$, and  (b) $\mathcal{O}$ remains non-zero at $N\rightarrow\infty$. In order to analyze the conditions (a) and (b) it is convenient to expand $\lvert E_0\rangle$ in the eigenbasis of the Hermitian operator $\hat{M}_{z}$. The operator $\hat{M}_{z}$ has $N+1$ distinct eigenvalues $m_k=(N-2k)/N$, where $k=0,1,...,N$ each having degeneracy $C^k_N$. Denoting the corresponding eigenstates as $\lvert m_k, d\rangle$, where $d=1,2,...,C^k_N$ is accounting for degeneracy, we can express $\lvert E_0\rangle=\sum_{k,d} c_{k,d} \lvert m_k,d\rangle$, where $c_{k,d}$ are normalized to unity, giving $\langle E_0\lvert \hat{M}_{z}\rvert E_0\rangle=\sum_{k,d} |c_{k,d}|^2 m_k$ and $\langle E_0\lvert \hat{M}_{z}^2\rvert E_0\rangle=\sum_k |c_{k,d}|^2 m_k^2$. The eigenbasis $\lvert m_k, d\rangle$ corresponds to states having $k$ spins pointing down and $N-k$ pointing up, and vice versa if we replace $k$ with $N-k$ ($m_k=-m_{N-k}$). Thus, in order to fulfill condition (a)  we can choose any superposition such that $|c_{k,d}|=|c_{N-k,d}|$.
%
%

One of the examples of the states satisfying the requirements discussed above, is the pair of maximally entangled Greenbergen-Horne-Zeilinger (GHZ) states \cite{Toth2005}
\begin{equation}
    \lvert G_{\mp}\rangle=\frac{1}{\sqrt{2}}(\lvert \uparrow\uparrow...\uparrow\rangle \mp \lvert \downarrow\downarrow...\downarrow\rangle),
\end{equation}
as they fulfill $\hat{M}_z\rvert G_{-,+} \rangle=\rvert G_{+,-} \rangle$ and $\langle G_{-,+}\lvert G_{+,-}\rangle=0$. Thus, the simplest possible TC Hamiltonian can be constructed as
\begin{align}\label{eq:simple_proj_ham}
    \hat{H}=-\lvert G_{+}\rangle\langle G_{+}\rvert,
\end{align}
where, in accordance with the notation introduced earlier, $\epsilon_0=-1$, $\epsilon_1=\epsilon_2=...=\epsilon_{2^N-1}=0$. 

If the ground state is $m$-degenerate, $m$ remains finite at $N\rightarrow\infty$, and for each $i=1,...,m$ ground state $\lvert E_0^{\{i\}} \rangle$ the quantity $\langle E_0^{\{i\}}\rvert \hat{M}_{z}^2\lvert E_0^{\{i\}}\rangle$ does not vanish in the thermodynamic limit, then the system remains a quantum time crystal for the averaging in Eq.~(\ref{eq:WO_definition}) performed over the zero temperature density matrix
\begin{align}
    \hat{\rho}=\frac{1}{m}\sum\limits_{i=1}^m\lvert E_0^{\{i\}}\rangle\langle E_0^{\{i\}}\rvert.
\end{align}
Moreover, the same conclusion holds for non-zero temperature once the contribution of excited states $~|\langle E_j |\hat{M}_z| E_k \rangle|^2$ remain finite in $N\rightarrow \infty$ limit. 

Generalizing the zero temperature case considered above, we proceed to the case of non-zero temperature $T=(k\beta)^{-1}$, when the system is described by the density matrix $\hat{\rho}=Z^{-1}e^{-\beta \hat{H}}$, where $Z=\sum_i e^{-\beta \epsilon_i}$. Without loss of generality, we focus on the system described by the Hamiltonian~(\ref{eq:ham_gen}) having the non-degenerate ground state. Then, the correlation function~(\ref{eq:WO_definition}) for finite $N$ reads 
\begin{align}
    f_N(t)=
    &Z^{-1}e^{-\beta \epsilon_0}\langle E_0\rvert \hat{M}_{z}^2\lvert E_0\rangle e^{i (\epsilon_0-\epsilon_1) t}\nonumber\\
    &+Z^{-1}\sum\limits_{j>0} e^{-\beta \epsilon_j} e^{i \epsilon_j t}\langle E_j\rvert \hat{M}_{z} e^{-i\hat{H}t}\hat{M}_{z}\rvert E_j \rangle,
\end{align}
and inserting the unity decomposition in the last line right after $e^{-i\hat{H}t}$ we obtain
\begin{align}
    f_N(t)=
    &Z^{-1}e^{-\beta \epsilon_0}\langle E_0\rvert \hat{M}_{z}^2\lvert E_0\rangle e^{i (\epsilon_0-\epsilon_1) t}\nonumber\\
    &+Z^{-1}\sum\limits_{j>0}\sum\limits_{k\ge 0} e^{-\beta \epsilon_j} e^{i (\epsilon_j-\epsilon_k) t}|\langle E_j\rvert \hat{M}_{z}\rvert E_k \rangle|^2.
\label{eq:fN_beta}
\end{align}
From the analysis of Eq.~\eqref{eq:fN_beta} we conclude that if there is a finite number of non-zero terms $\langle E_j\rvert \hat{M}_{z}\rvert E_k \rangle$ in the thermodynamic limit and $Z$ does not diverge in the same limit, then the system represents a quantum time crystal for any temperature.

\textit{Locality of TC Hamiltonian.---}Above we have presented the quantum time crystal Hamiltonian by construction, formulated in terms of the many-body projector and coupling operators. However, when it is rewritten in terms of physical interaction between qubits [e.g. taking the toy TC Hamiltonian from \eqref{eq:simple_proj_ham}], it reveals highly nonlocal character. One can easily check that resulting expression contains $N$-length spin strings of $\sigma_x$ operators with pairwise substitution of $\sigma_y^{(i)} \sigma_y^{(j)}$ couples, and even-length strings of $\sigma_z$ operators \cite{NoteBehemoths}. This corresponds to the sum of GHZ-state stabilizer products \cite{Toth2005}, which must satisfy the conditions specified for Hamiltonian \eqref{eq:ham_gen}. While there are various ways the spin-strings can be combined, one can consider a particular example and verify that W-O conditions for TC are satisfied. For instance, taking the Hamiltonian of an $N$-qubit system in the form
\begin{equation}\label{eq:H_TC_xy}
    \hat{H}=-\frac{J}{N(N-1)}\sum\limits_{1\le i<j\le N}\sigma^{(1)}_x\sigma^{(2)}_x...\sigma^{(i)}_y...\sigma^{(j)}_y...\sigma^{(N)}_x,
\end{equation}
where there are exactly two $\sigma_y$ operators in each term filled with $\sigma_x$'s, we observe that its spectrum is bounded between $-J/2\le \epsilon_n\le J/2$. The non-degenerate ground state of the Hamiltonian~\eqref{eq:H_TC_xy} is given by $\lvert G_{-}\rangle$ with energy $\epsilon_0=-J/2$, and the highest exited state corresponding to $\epsilon_{2^N-1}=J/2$ energy (which is also non-degenerate) reads $\lvert G_{+}\rangle$. Thus, the Hamiltonian~(\ref{eq:H_TC_xy}) fulfills W-O TC definition~\cite{Oshikawa} in the thermodynamic limit. It is worth noting that one could take a correlation function of the form $\langle E_0\rvert \sigma^{(i)}_z(t)\sigma^{(j)}_z(0)\rvert E_0 \rangle$ and get absolutely the same result for the correlation function~(\ref{eq:WO_definition}), namely $e^{-iJt}$ (the two correlation functions coincide for an arbitrary $N$). Thus, the provided Hamiltonian \eqref{eq:H_TC_xy} implements a genuine time crystal formally bypassing W-O no-go theorem by exploiting nonlocality. We also note that long-range multispin interaction Hamiltonians may be potentially studied in the systems with slow but strong driving, where heating is precluded \cite{Haldar2018,Lazarides2014}.
\begin{figure}
\includegraphics[width=1.\columnwidth]{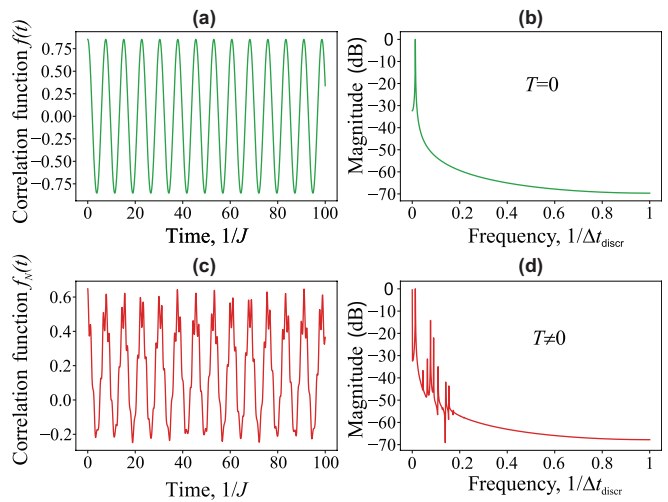}
\caption{Plot (a) shows the correlation function corresponding to TC (time crystal) behavior at zero temperature $T=0$ for an arbitrary size of the system $N$ and (b) shows its spectrum ($\Delta t_{\text{discr}}$ is the discretization step in time). The deviation from the ideal TC behavior at $T\neq0$ [$\beta=(kT)^{-1}=1$] is shown in the panel (c) for the finite system of $N=10$ sites, and its spectrum is in (d). In this case, the TC behavior is expected to vanish in the thermodynamic limit, when the number of additional harmonics tends to infinity, ruining the periodicity of the correlation function, whereas for the finite sized system the correlation function remains periodic with large but finite period. }
\label{fig:dynamics}
\end{figure}

Finally, we can ask the question: what is the most local Hamiltonian which satisfies TC conditions described after Eq.~\eqref{eq:ham_gen}? To answer it, one needs to find the Hamiltonian containing in the spectrum $\lvert G_{+}\rangle$ and the shortest interaction range. The ultimate bound on such interaction was given in Ref.~\cite{FlorioPRL2011}, where the most local Hamiltonian comprises of at least $[N/2]$-length strings. It can be written as $\hat{H}(J)=\hat{H}_0 + J\hat{H}_1$, and contains the nearest-neighbour Ising Hamiltonian 
\begin{equation}\label{eq:Ising}
    \hat{H}_0 = -\sum\limits_{j=1}^N\sigma_z^{(j)}\sigma_z^{(j+1)},
\end{equation}
for the periodic boundary conditions $(\sigma_z^{(N+1)}\equiv\sigma_z^{(1)})$, and spin-string interaction part
\begin{equation}
    \hat{H}_1=\sigma_x^{(1)}\sigma_x^{(2)}...\sigma_x^{([N/2])}-\sigma_x^{([N/2]+1)}...\sigma_x^{(N)},
\end{equation}
which involves the half of the qubits. The ground state $\lvert \mathrm{GS}\rangle$ of the Hamiltonian $\hat{H}(J)$ satisfies $M_z\lvert \mathrm{GS}\rangle\sim\lvert G_{+}\rangle$  and $\langle \mathrm{GS}\lvert M_z^2\rvert \mathrm{GS}\rangle$ is finite at $N\rightarrow\infty$. These states have the energies $\hat{H}(J)\lvert G_{+}\rangle=-N\lvert G_{+}\rangle$ and $\hat{H}(J)\lvert \mathrm{GS}\rangle=[-N-2(\sqrt{1+J^2}-1)]\lvert \mathrm{GS}\rangle$.
The correlation function~(\ref{eq:WO_definition}) for the Hamiltonian $\hat{H}(J)$ is given by
\begin{align}
&\lim_{N\rightarrow\infty}\langle \mathrm{GS}\rvert e^{i\hat{H}(J)t}\hat{M}_{z} e^{-i\hat{H}(J)t}\hat{M}_{z}\lvert \mathrm{GS}\rangle\nonumber\\
&=\lim_{N\rightarrow\infty}\langle \mathrm{GS}\lvert M_z^2\rvert \mathrm{GS}\rangle e^{-i2(\sqrt{1+J^2}-1)t}\nonumber\\
&=\sin^2{(\frac{3\pi}{8})}e^{-i2(\sqrt{1+J^2}-1)t},    
\end{align}
thus corresponding to the time crystalline behavior for the most local TC Hamiltonian $\hat{H}(J)$. In the absence of $\hat{H}_1$ the Hamiltonian~$\hat{H}(J)$ reduces to the Ising Hamiltonian having $|G_{+,-}\rangle$ as degenerate ground states. 

\textit{Stability of time crystal.---}The important feature of time crystal phase is its stability towards perturbations in the Hamiltonian (extra magnetic field), which in the case of DTC has allowed to distinguish from otherwise trivial spin flip operation \cite{Else2016,Yao2017}. Examining the behavior of the Hamiltonian $\hat{H}(J)$ under magnetic field perturbations, $\hat{H}' = \hat{H}(J) + \delta \hat{H}$ with $\delta \hat{H}=\sum_j (h_x^j \sigma_x^{(j)}+h_y^j \sigma_y^{(j)}+h_z^j \sigma_z^{(j)})$, we note that $\langle \mathrm{GS}\rvert \delta\hat{H}\lvert \mathrm{GS}\rangle=0$ and $\langle G_{+}\rvert \delta\hat{H}\lvert G_{+}\rangle=0$. Moreover, the perturbation of the form $\delta \hat{H}=\sum_j \sigma_{x,y}^{(j)}\sigma_{x,y}^{(j+1)}$ analogously does not overlap with $\lvert \mathrm{GS}\rangle$ or $\lvert G_{+}\rangle$. Thus, these types of perturbations do not affect the part of the Hamiltonian that is responsible for TC behavior, $\hat{H}_{01} = \epsilon_0\lvert E_0\rangle\langle E_0\rvert+ \epsilon_1\lvert E_1\rangle\langle E_1\rvert$ (for $\hat{H}(J)$ the ground state is $\lvert E_0\rangle\equiv\lvert \mathrm{GS}\rangle$ and $\lvert E_1\rangle\sim\lvert G_{+}\rangle$).

If the temperature is non-zero, then the analysis of stability becomes more subtle as exited states come into play and contribute to the correlation function, leading to the generation of higher harmonics. Therefore, it is important to understand the scaling of the number of additional distinct harmonics with the size of the system. At non-zero temperature, our investigation of the Hamiltonian $\hat{H}(J)$ is limited to numerical analysis based on exact diagonalization which showed that the number of additional distinct harmonics grows with the size of the system. Thus, we expect the melting of the time crystal at non-zero temperature, which is shown in~Fig.~\ref{fig:dynamics}. However, the analysis of the TC phase at non-zero temperature in the thermodynamic limit remains an open question. 

\begin{figure}
\includegraphics[width=1.\columnwidth]{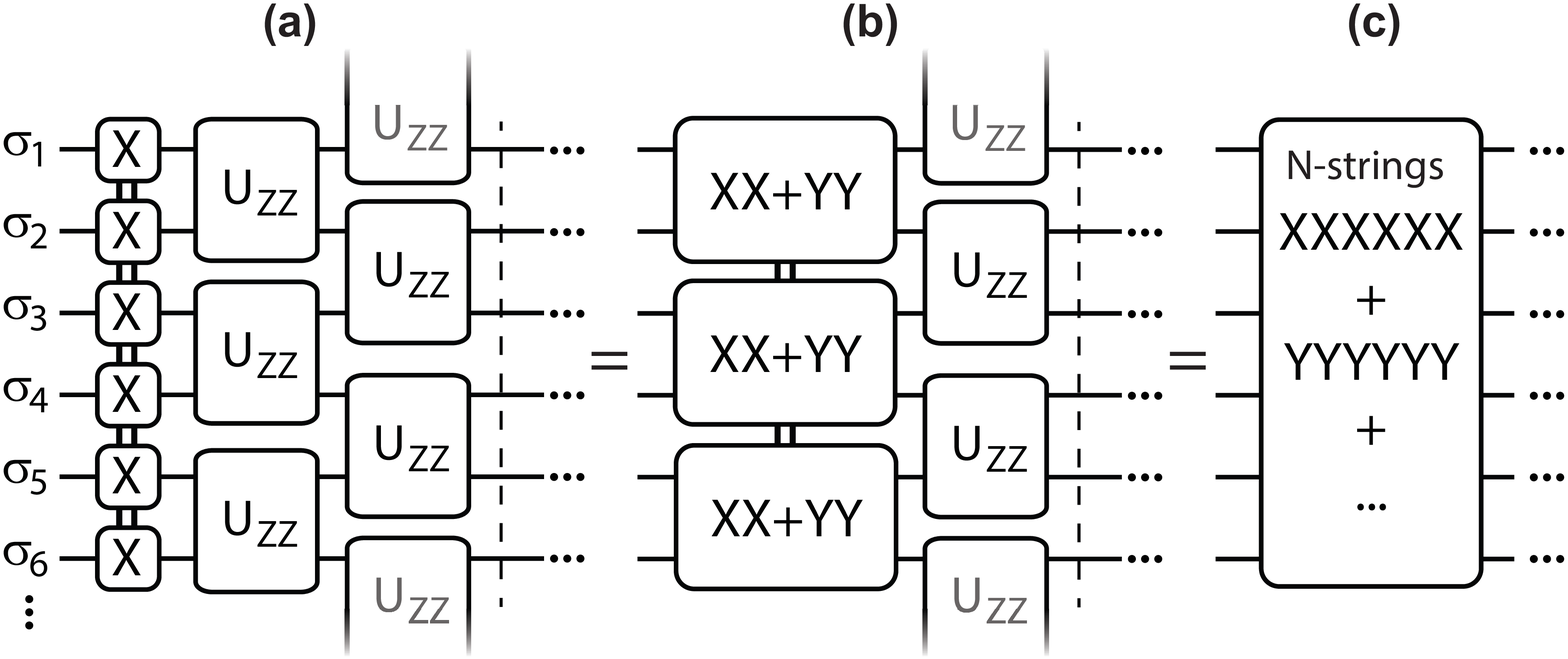}
\caption{The protocol for Floquet time crystal generation, and associated reasoning for existence of continuous time crystals with static $\hat{\mathcal{H}}_{\mathrm{TC}}$. The full digital sequence is presented in (a), which consists of $\pi$ rotations around $\mathbf{x}$ axis, corresponding to spin flip operators ($\mathrm{X}$) forming a string. The unitary corresponding to Ising interaction is divided into $U_{\mathrm{ZZ}}$ commuting pieces. Next, $\mathrm{X}$ string and odd sublattice Ising terms are combined, leading to the product of $\mathrm{XX+YY} := -\cos(\phi_j/2) \sigma_x^{(j)} \sigma_x^{(j+1)} - \sin(\phi_j/2) \sigma_y^{(j)} \sigma_y^{(j+1)}$ operators (b). Finally, during the last step all layers are linked together, forming the string (c).}
\label{fig:sketch}
\end{figure}

\textit{Relation to discrete time crystals.---}To better understand the origin of TC behavior for continuous time, we can study its relation to the discrete counterpart \cite{Else2016}. We recall that standard DTC protocol is based on the unitary operation comprising a sequence of gates for an $N$-qubit chain. It reads as: 1) start with all-down or all-up spin string and apply spin flip operator $\hat{U}_{\mathrm{X}} = \exp(-i \pi/2 \sum_j \sigma_x^{(j)}) = \prod_j (-i \sigma_x^{(j)})$; 2) evolve the state with Ising Hamiltonian $\hat{U}_{\mathrm{Ising}} = \exp(-i \tau \sum_{j=1}^{N} J_{j,j+1} \sigma_z^{(j)} \sigma_z^{(j+1)})$, thus completing the single step unitary $\hat{\mathcal{U}}_{\mathrm{step}} = \hat{U}_{\mathrm{Ising}} \cdot \hat{U}_{\mathrm{X}}$; 3) continue for many stroboscopic periods $\hat{\mathcal{U}}_{\mathrm{DTC}} = \hat{\mathcal{U}}_{\mathrm{step}} \cdot \hat{\mathcal{U}}_{\mathrm{step}} \cdot ...$~. The algorithm is sketched in Fig. \ref{fig:sketch} (a). The resulting stroboscopic dynamics then shows persistent oscillations of $M_z$ magnetization with period twice bigger than the drive $\prod_j (-i \sigma_x^{(j)})$.
Given the unitary operation, we can formally define the static Hamiltonian which governs the dynamics as $\hat{H}_{\mathrm{DTC}} = i \log (\hat{\mathcal{U}}_{\mathrm{step}})$. Generally, the logarithm can be computed the using Dynkin's formula, but the infinite order resummation needed to obtain the closed form expression seems infeasible. Instead, we can analyze the structure of $\hat{H}_{\mathrm{DTC}}$ by studying the unitary operator in Fig. \ref{fig:sketch} (a). First,  exploiting the Euler's formula we can combine the $\sigma_{x}$-string and the layer of Ising terms into a single block. Denoting phases as  $J_{j,j+1} \tau = \phi_j$, the odd $j$ sublattice unitary reads $\prod_j (-1) \{ \cos(\phi_j/2) \sigma^{(j)}_x \sigma^{(j+1)}_x + \sin(\phi_j/2) \sigma^{(j)}_y \sigma^{(j+1)}_y \}$, which we denote as blocks ``XX + YY'' in Fig. \ref{fig:sketch} (b). These blocks multiplied altogether contain various $\sigma_x$ and $\sigma_y$ strings, from length $2$ to length $N$. Finally, the multiplication of the second sublattice Ising interaction either leaves string unaltered, or changes $\sigma_x \leftrightarrow \sigma_y$ (with some phase factor). Once the matrix $\hat{\mathcal{U}}_{\mathrm{step}}$ is obtained, the effective Hamiltonian can be inferred employing the replica trick \cite{Vajna2017}. This relies on taking formally the limit
\begin{align}
\label{eq:replica}
\log \hat{\mathcal{U}}_{\mathrm{step}} = \lim_{\rho \rightarrow 0} \frac{1}{\rho} \Big(\hat{\mathcal{U}}_{\mathrm{step}}^{\rho} - 1 \Big),
\end{align}
and corresponds to taking the powers of previously derived unitary matrix. In this case, the odd powers contribute to the same stings we observe in the Hamiltonian \eqref{eq:H_TC_xy} [see Fig.~\ref{fig:sketch}(c)], albeit with different prefactors, and cross-terms from even powers contribute to the $\sigma_z$-strings of even length. 



Finally, let us relate the physics of discrete time crystals through the properties of $\hat{H}_{\mathrm{DTC}}$ Hamiltonian and continuous time crystal Hamiltonian described in the previous sections. In particular, if the ground state for both Hamiltonians is all-down or all-up spin string or their superposition, than one may expect that breaking of continuous TTS is inherently related to breaking of the discrete TTS for the corresponding Floquet TC. To yield direct comparison, we can consider the odd size $N$ system with Ising interaction fixed $\phi_{j,j+1} = J_{j,j+1} \tau=-1/N$. Then, the matrix exponentials and logarithm can be calculated numerically, and their analysis provides a concise structure of the effective DTC Hamiltonian
%
\begin{align}
\label{eq:TC_ham_from_DTC}
    \hat{H}_{\mathrm{DTC}}=-\frac{1}{N}\sum\limits_{j=1}^{N}\sigma_z^{(j)}\sigma_z^{(j+1)}+(-1)^{(N-1)/2}\frac{\pi}{2}\prod_j\sigma_x^{(j)},
\end{align}
which is analogous to the time crystal Hamiltonian $\hat{H}(J)$ introduced earlier in a sense that it has a form of the Ising Hamiltonian plus a non-local part lifting the degeneracy of the ground state. For $N=4k+3$ the non-degenerate ground state of the Hamiltonian~(\ref{eq:TC_ham_from_DTC}) is $\lvert G_{+}\rangle$ and $\lvert G_{-}\rangle$ is an exited state, whereas for $N=4k+1$ it is the other way round. The energy difference between $\lvert G_{\pm}\rangle$ states is equal to $\pi$ for any odd $N$. Therefore, the Hamiltonian~(\ref{eq:TC_ham_from_DTC}) is nothing but a quantum time crystal and the corresponding correlation function for any odd $N$ is $f(t)=e^{-i\pi t}$.

\textit{Conclusions and outlook.---}While previously it was believed to be impossible in the case of non-driven system, we have shown that time crystals can emerge for Hamiltonians with long-range interactions in the form of spin strings. We described the minimal continuous TC Hamiltonian as a sum of strings encompassing halves of a spin system, and find that at zero temperature TC is stable. Finally, we note that nonlocal many-body operators which induce time crystalline behavior represent a specific example of hermitian quantum behemoths~\cite{Khaymovich2019}, possibly relating the two notions.

\begin{acknowledgments}
\textit{Acknowledgments.---}The work was supported by the Government of the Russian Federation through the ITMO Fellowship and Professorship Program, Megagrant 14.Y26.31.0015, and Goszadanie no. 3.2614.2017/4.6 and 3.1365.2017/4.6 of the Ministry of Education and Science of Russian Federation. The authors acknowledge the help from NORDITA Visiting PhD program. We are grateful to Frank Wilczek and Anatoli Polkovnikov for useful discussions. We also would like to thank Ivan Iorsh for the fruitful suggestions.
\end{acknowledgments}



\begin{thebibliography}{99}
\bibitem{NoteBehemoths} The origin lies in a non-trivial operator representation of many-body projectors, which can be checked by direct decomposition.
\end{thebibliography}
\end{document}